# The "Alternative" Choice of Constitutive Exons throughout Evolution

Galit Lev-Maor[1]●, Amir Goren[1]●, Noa Sela[1]●, Eddo Kim[1], Hadas Keren[1], Adi Doron-Faigenboim[2], Shelly Leibman-Barak[3], Tal Pupko[2], Gil Ast[1]*

1 Department of Human Molecular Genetics, Tel Aviv University, Tel Aviv, Israel, 2 Department of Cell Research and Immunology, Tel Aviv University, Tel Aviv, Israel, 3 Department of Neurobiochemistry, Tel Aviv University, Tel Aviv, Israel

Alternative cassette exons are known to originate from two processes—exonization of intronic sequences and exon shuffling. Herein, we suggest an additional mechanism by which constitutively spliced exons become alternative cassette exons during evolution. We compiled a dataset of orthologous exons from human and mouse that are constitutively spliced in one species but alternatively spliced in the other. Examination of these exons suggests that the common ancestors were constitutively spliced. We show that relaxation of the 5′ splice site during evolution is one of the molecular mechanisms by which exons shift from constitutive to alternative splicing. This shift is associated with the fixation of exonic splicing regulatory sequences (ESRs) that are essential for exon definition and control the inclusion level only after the transition to alternative splicing. The effect of each ESR on splicing and the combinatorial effects between two ESRs are conserved from fish to human. Our results uncover an evolutionary pathway that increases transcriptome diversity by shifting exons from constitutive to alternative splicing.

## Introduction

Splicing is a mechanism by which mature mRNA is formed through the removal of introns from the pre-mRNA and the ligation of exons. Four short sequences define an intron: the 5′ and 3′ splice sites (5′ss and 3′ss), the branch site (BS), and the polypyrimidine tract (PPT) [1]. Alternative splicing is a process by which more than one mRNA is produced from the same pre-mRNA [2,3]. Alternative splicing events are classified into four main subgroups: (i) removal of alternative cassette exons (also called exon skipping), when an exon and flanking introns are spliced out of the transcript; (ii) use of alternative 5′ss or (iii) 3′ss, resulting from recognition of two or more splice sites at one end of an exon; and (iv) intron retention, in which an intron remains in the mature mRNA molecule [4,5]. In mammals, a large fraction (40%–80%) of the genes undergoes alternative splicing, but no alternative splicing has been reported in *Saccharomyces cerevisiae* or *S. pombe* [6]. Thus, alternative splicing is believed to be a major source of the phenotypic complexity in higher eukaryotes [2,7]. The recent availability of genomes and transcriptomes provides data for the study of the evolution of alternative splicing.

Several large-scale bioinformatic studies have examined the evolution of alternative splicing. One of the primary works demonstrated that alternative cassette exons with a high inclusion level are usually conserved between human and mouse (i.e., they are found in both genomes), but that alternative cassette exons with a low inclusion level are often not conserved [8]. Additionally, these low-inclusion alternative cassette exons most likely originated from unique intronic sequences [9]. Previous studies demonstrated that homologous exons that are spliced differently in human and mouse, termed species-specific exons, represent at least 11% of the human–mouse alternative cassette exons and exhibit a high inclusion level, presumably reflecting their evolutionary history. These exons probably have an important role in the evolutionary differences among mammalian species [10,11]. These results are indicative of the dynamic evolution of alternative cassette exons and their major contribution to genome structure and transcriptomic diversity?

But how did alternative splicing evolve? One known mechanism for generating new alternative cassette exons is exon shuffling, in which a new exon is inserted into an existing gene or an exon is duplicated within the same gene [12–14]. Alternative cassette exons can also emerge following exonization of intronic sequences; about 4%–5% of human genes contain transposable element motifs in their coding regions [15–19]. These newly inserted exons typically exhibit a low inclusion level, thus maintaining the evolutionary conserved isoform as the main mRNA product [6].

In the two mechanisms described above, the alternative cassette exons are added to existing genes. However, a third mechanism was recently proposed, in which alternative exons are derived from constitutively spliced ones [6]. Recently, we demonstrated that alternative 3′ and 5′ exons can originate

Abbreviations: 3′ss, 3′ splice sites; 5′ss, 5′ splice sites; AS exon, alternatively spliced exon; BS, branch site; CDS, coding sequence; ESE, exonic splicing enhancer; ESR, exonic splicing regulatory sequences; ESS, exonic splicing silencer; EST, expressed sequence tags; HI, high inclusion group; Ka, nonsynonymous substitution rate; Ka/Ks, ratio between nonsynonymous and synonymous substitution rate; Ks, synonymous substitution rate; LI, low inclusion group; PPT, polypyrimidine tract; RT-PCR, Reverse transcription-PCR; WT, wild type

* To whom correspondence should be addressed. E-mail: gilast@post.tau.ac.il

● These authors contributed equally to this work.

## Author Summary

Alternative splicing is believed to play a major role in the creation of transcriptomic diversification leading to higher order of organismal complexity, especially in mammals. As much as 80% of human genes generate more than one type of mRNA by alternative splicing. Thus, alternative splicing can bridge the low number of protein coding genes (~24,500) and the total number of proteins generated in the human proteome (~90,000). The correlation between the higher order of phenotypic diversity and alternative splicing was recently demonstrated and thus the origin of alternative splicing is of great interest. There are currently two models regarding the origin of alternatively spliced exons—exonization of intronic sequences and exon shuffling. According to these two mechanisms, a protein-coding gene was first established and only then a new alternative exon appeared within it or was added to the gene. Our current study provides evidences for a new mechanism indicating that during evolution constitutively spliced exons became alternatively spliced. Large-scale bioinformatic analyses reveal the magnitude of this process and experimental validation systems provide insights into its mechanisms.

from constitutively spliced exons by the generation of a competing splice site, either within the exon or in one of the flanking introns, thus generating new exons from constitutive exons [20]. Here, we provide computational and experimental evidence supporting the hypothesis that alternative cassette exons can also originate from constitutively spliced exons. A dataset of human–mouse exons that are spliced in a species-specific manner was compiled. Bioinformatic analysis of these exons using expressed sequence tags (ESTs) from mammalian and nonmammalian organisms suggests that the common ancestors of species-specific alternatively spliced exons (AS exons) were constitutively spliced exons. The reliability of the EST analysis was confirmed experimentally. The alternatively spliced, species-specific AS exons had a high inclusion level, presumably to ensure synthesis of the ancient spliced form (exon inclusion) as the major mRNA product. Relaxation of 5'ss selection was shown to be involved in the evolutionary shift from constitutive to alternative splicing. The transition to alternative splicing was associated with fixation of exonic splicing regulatory sequences (ESRs) that control the inclusion/skipping ratio in alternative splicing. The effect of each ESR on splicing and the combinatorial effects between two ESRs are conserved among zebrafish, chicken, mouse and human. Our results support the existence of an evolutionary pathway that increases transcriptome diversity by shifting exons from constitutive to alternative splicing.

## Results

### Conserved Species-Specific Exons Contain Characteristics of Constitutive Exons

We compiled three datasets containing orthologous exons from human and mouse and their flanking introns [21]: exons constitutively spliced in both species (45,553), alternative cassette exons in both species (596), and species-specific exons (612). In the species-specific exon dataset, 354 and 258 exons are alternatively spliced in human and mouse, respectively.

Certain characteristics distinguish alternatively spliced from constitutively spliced exons [11,22–25]. Based on these characteristics, we performed a large scale analysis using 15 parameters to test whether species-specific AS exons were more similar to constitutively spliced exons or alternative cassette exons (Figure 1A). Our results showed that in 14 out of the 15 tested characteristics, the species-specific AS exons were more similar to the constitutively spliced exons than to the alternative cassette exons (Figure 1A and Table S1). For example, constitutively spliced exons exhibited a high degree of exon sequence conservation (termed the identity level) of 87.8%, whereas alternative cassette exons that exhibited a 92.2% identity level. The average conservation level between species of species-specific exons was 88.3%, similar to that of constitutively spliced exons. Although the species-specific AS exons differed significantly from constitutively spliced exons in the tested characteristics (Mann-Whitney, $p$ value = 0.04), the difference was much more significant when compared with alternative cassette exons (Mann-Whitney, $p$ value = $6 \times 10^{-34}$; see Table S1 for all tested parameters and $p$ values).

A ratio between nonsynonymous and synonymous substitution rates (Ka/Ks ratio) test was previously proposed for detecting exons in genomic regions [26]. Xing and Lee (2005) showed that this method efficiently detects exons that are constitutively spliced but is less efficient in the case of alternative cassette exons [27]. This observation suggests that this ratio test can also be used to distinguish constitutively spliced exons from alternatively spliced ones. We note that this test differs from the Ka/Ks ratio, as it takes into account the standard error in the Ka/Ks estimates (i.e., it statistically tests whether the exon evolves). As expected, constitutively spliced exons were found to be significantly different from alternatively spliced ones: the percentage of exons that failed the Ka/Ks ratio test was 7.6% and 33.3% for constitutively and alternative cassette exons, respectively (Fisher Exact, $p$ value < $4.2 \times 10^{-50}$). In agreement with our results above, the Ka/Ks ratio test on the species-specific AS exons resulted in 13.5% failure, indicating a high similarity to constitutively spliced exons. Statistically, the AS exons differed significantly from constitutive exons (Fisher Exact, $p$ value = $1.8 \times 10^{-6}$). However, the difference was much larger when compared to alternative exons (Fisher Exact, $p$ value = $2.2 \times 10^{-13}$) (Figure 1A).

Since we had established that the species-specific AS exons were more similar to constitutive exons than to alternative cassette exons, we next tested whether the flanking intronic sequences also showed this tendency. We demonstrated previously that intronic sequences flanking alternative cassette exons conserved in human and mouse are more conserved than constitutively spliced exons [28]. We found that an average of 119.4 nucleotides upstream and 117.4 nucleotides downstream of exons could be aligned in constitutively spliced exons, compared with 233.4 and 224.7 in alternative cassette exons ($p$ value = 0 and $p$ value = 0 for upstream and downstream introns, Mann-Whitney test). The intronic sequence alignment of the species-specific AS exons are very similar to the results of constitutively spliced exons: 125.5 and 116.6 nucleotides could be reliably aligned upstream and downstream of exons, respectively (Figure 1A and 1B). The species-specific AS exons did not statistically differ from constitutively spliced exons ($p$ value = 0.085 and $p$ value = 0.39 for upstream and downstream introns, Mann-Whitney test), but significantly differed from alternative cassette exons ($p$ value = $2.58 \times 10^{-52}$ and $p$ value = $1.47 \times 10^{-58}$ for upstream and downstream introns, Mann-Whitney test).

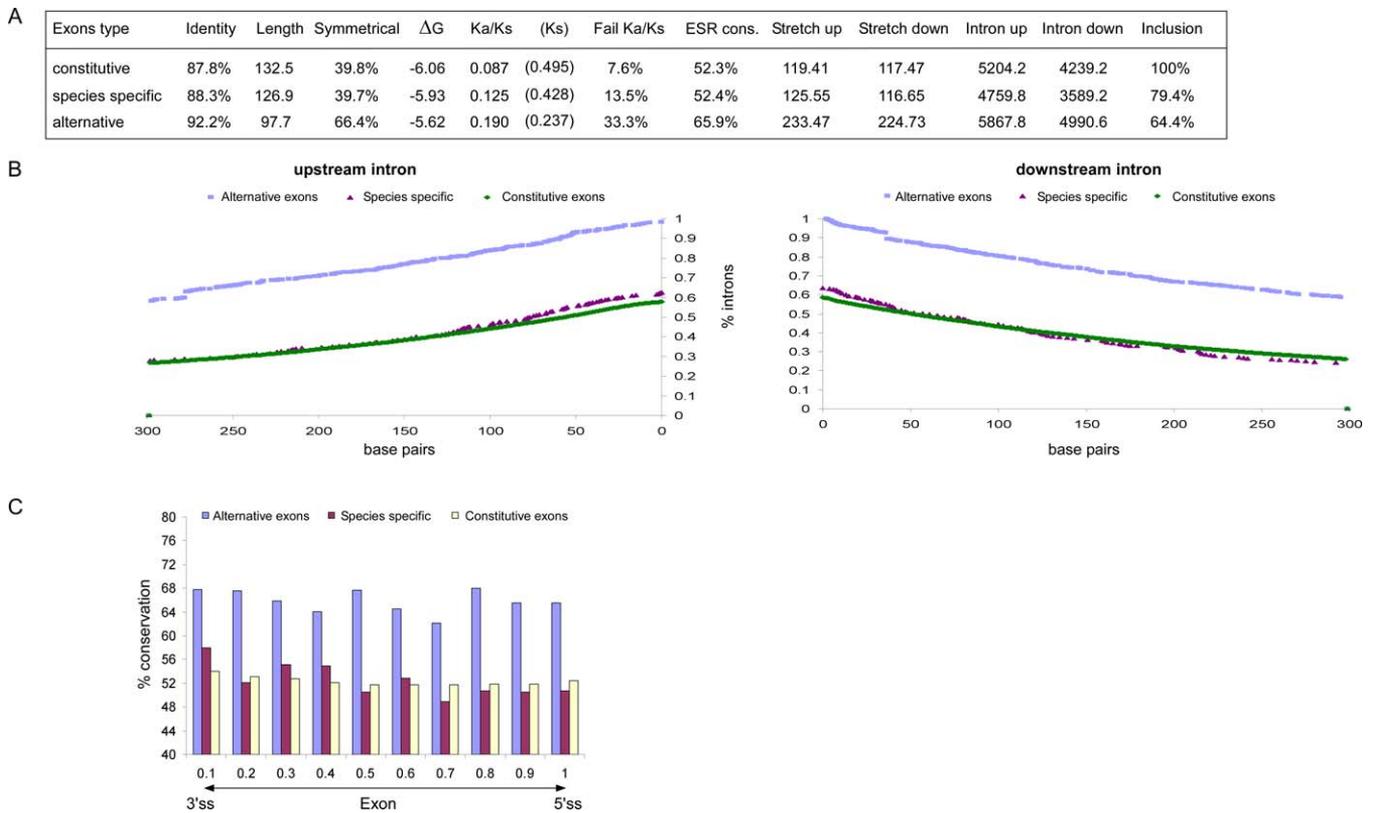

**Figure 1.** Species-Specific Exons Exhibit Characteristics of Constitutive Exons

(A) Fifteen different characteristics that differentiate constitutively spliced exons from alternative cassette exons were examined on orthologous human and mouse exons that are constitutively spliced in one species but alternatively spliced in the other. We examined exon sequence conservation level between human and mouse (Identity), exon lengths (Length), percentage of exons in which the total number of nucleotides was divisible by 3 (Symmetrical), U1/5'ss binding strength ($\Delta G$), the ratio between nonsynonymous and synonymous substitution rates (Ka/Ks), synonymous substitutions rate (Ks), the fraction of exons that fail the Ka/Ks likelihood ratio test (Fail Ka/Ks), the conservation of ESRs along the exons (ESR cons), the size of the flanking introns stretches that could be reliably aligned to the mouse ortholog for both upstream and downstream intron (Stretch up and Stretch down, respectively), the average size of the upstream and downstream introns (Intron up and Intron down, respectively), and the inclusion level of the exons (Inclusion). (B) Conservation level of the flanking intronic sequences. Median conservation level at each base position was calculated for the 300 base pairs upstream (left) and downstream (right) of the exon for constitutively spliced exons (green circles), alternative cassette exons (blue squares), and species-specific exons (red triangles). (C) The relative conservation level of ESRs between human and mouse, where the X axis is the location, from the 3'ss to the 5'ss in 10% increments (exon size is normalized between 0 and 1) and the Y axis is the conservation level of the wobble positions between conserved human–mouse orthologous exons that are constitutively (yellow) and alternatively (blue) spliced in both human and mouse, or species-specific exons (red).

doi:10.1371/journal.pgen.0030203.g001

ESRs are short exonic sequence motifs that regulate splicing. Conservation of these elements is different in constitutive and alternative cassette exons [29]. Thus, we compared the conservation level in human and mouse sequences of 285 putative ESRs in our three datasets (Figure 1C). As expected, ESRs were significantly more conserved in alternative cassette exons compared with constitutively spliced exons (65.9% and 52.3% conservation, respectively; $\chi^2$, $p$ value $= 9.6\times10^{-92}$). The species-specific AS exons exhibited a 52.4% conservation level of the ESRs, which is statistically insignificant with respect to constitutively spliced exons, but statistically significant with respect to alternative cassette exons ($\chi^2$, $p$ value $= 3.2\times10^{-56}$). Figure 1A–1C show that the species-specific AS exons are more closely related to constitutively spliced exons, in 14 out of 15 different characteristics, than to alternative cassette exons.

## High Inclusion Level of Species-Specific AS Exons

The alternative variant of species-specific spliced exons is either the ancestral form or the derived one. To distinguish between these two possibilities, the splicing pattern of exons that are orthologous to the species-specific AS exons was analyzed experimentally. We selected orthologous exons that are alternatively spliced in human and constitutively spliced in mouse, for which orthologous exons and the conserved flanking exons could be identified using the HomoloGene database for six different organisms—human, mouse, rat, dog, chicken, and zebrafish. Ten exons were initially evaluated (Table S2). In five cases, the orthologous exons and the conserved flanking exons were found in all six organisms. For each orthologous exon, pairs of primers were designed to hybridize in flanking exons to amplify a PCR product of similar size for the exon-inclusion isoform in all organisms. Total RNA was extracted from brain tissues of human, mouse, rat, and dog, and from the entire organism of chicken and zebrafish. The RNA was analyzed by reverse transcription-PCR (RT-PCR) and by sequencing of all PCR products. Similar results were obtained when RNA was extracted from different cell lines from human, mouse, and rat (see Text S2).

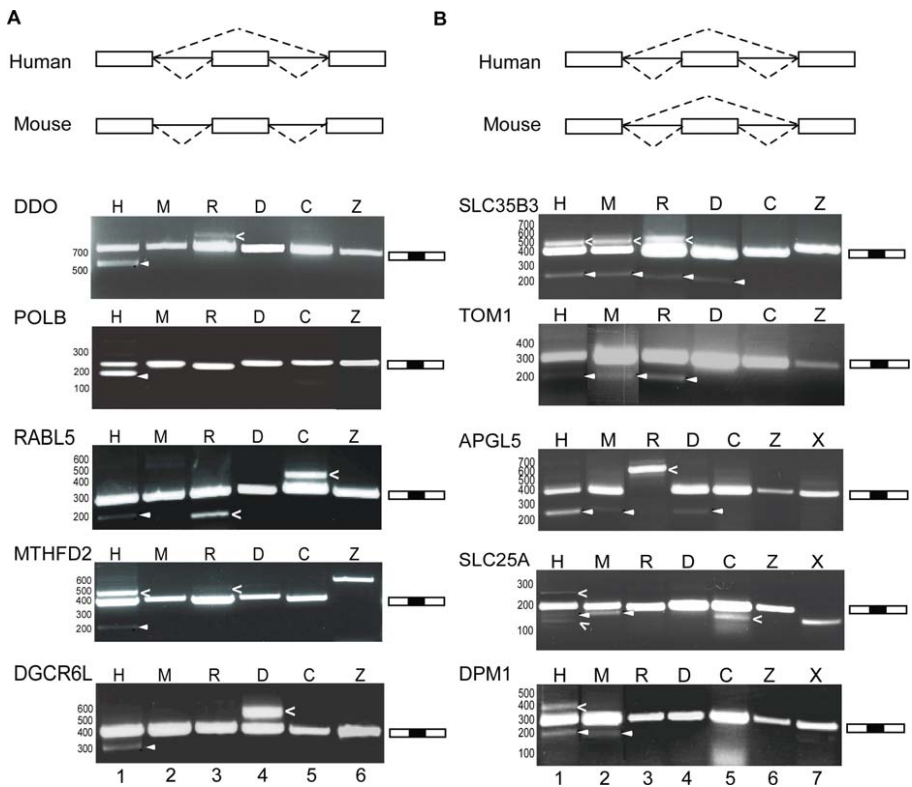

**Figure 2.** Changes in the Splicing Pattern of Orthologous Exons during Evolution

(A) Schematic illustration of the bioinformatics analysis in which the orthologous exon is alternative in human and constitutive in mouse (upper panel). The splicing pattern of five of those exons was tested in several organisms. RT-PCR analysis is shown for normal brain cDNA of human, mouse, rat, and dog (marked H, M, R, and D, respectively), chicken (5-day-old embryos, marked as C), and adult zebrafish whole body (Z). PCR products were amplified using species-specific primers and splicing products were sequenced on a 2% agarose gel and sequenced. (B) Similar analysis as in panel A, except that the orthologous exons are alternatively spliced with a high inclusion level in both human and mouse. cDNA from *Xenopus* oocytes (marked X) were also analyzed for some of the orthologs. For *APGL5* (lane R), *DPM1* (lane H upper band), *SLC25A* (lane H upper and lower bands) and lane C lower band), see Text S5. Gene names are shown on the left (see also Table S2). Arrowheads point to exon-skipping isoform; exon-inclusion isoforms are shown on the right by black and white boxes; open arrows point to non-conserved alternatively spliced products (see Text S5).
doi:10.1371/journal.pgen.0030203.g002

The experimental results validate that the exons tested are indeed species-specific: These exons are alternatively spliced in human and constitutively spliced in mouse (lanes H and M, for human and mouse, respectively, Figure 2A). Moreover, the alternative form was unique to human, as all other tested organisms exhibited a constitutive splicing pattern (Figure 2A, lanes R, D, C, and Z, for rat, dog, chicken, and zebrafish, respectively). Additional nonconserved spliced products were observed (marked by open arrows), such as intron retention and alternative 3'ss usage (all PCR products were sequenced; see Text S5 for their identity). Based on the phylogenetic relationships among the analyzed organisms, we conclude that the alternative splice variant is a derived form and the constitutively spliced variant is the ancestral one.

In human, exon inclusion was the major splicing product, and use of alternative cassette exons is the minor one for all genes analyzed except for *POLB*, in which the exon inclusion/ skipping ratio was about 50% (Figure 2A). The high inclusion level was consistent with the bioinformatic analysis (which is one of the characteristics in Figure 1A) and with a previous report [10]. The inclusion level analysis suggests that the evolutionary shift from constitutive to alternative splicing was from fully constitutively spliced exons to alternatively spliced exons with a high inclusion level. We note that equivalent amounts of cDNA were used in all PCR reactions (Figure 2) and, therefore, the amounts of PCR products presumably reflect different levels of mRNA (for example, compare lane Z, Figure 2, between *DGCR6L* and *TOM1*). Moreover, whenever the alternative cassette exon's isoforms were not detected, longer exposure of the gels did not indicate their presence. Some of the alternative cassette exons events generate an isoform with a premature stop codon and we have shown that low RNA levels in human cell lines are not due to degradation through the nonsense mediated mRNA decay (NMD) pathway (unpublished data). These results suggest that many species-specific AS exons originate from ancestral constitutively spliced exons.

## Exons with High Inclusion Levels Also Originate from Constitutively Spliced Exons

To further examine the hypothesis that alternative exons originate from ancestral constitutively spliced exons, we selected five alternative cassette exons with a high inclusion level conserved in human and mouse and tested their splicing patterns in different organisms by RT-PCR analysis (Figure 2B). The relevant exon in the *SLC35B3* gene was alternatively spliced in all four mammals with a high inclusion level, but constitutively spliced in chicken and zebrafish. Additionally, in human, mouse, and rat there is apparently an exon repetition isoform (marked by open arrows, see Supplemental Materials, and also [30]). The alternatively spliced exon in the *TOM1* gene was detected in the Euarchontoglires (the phylogenetic group that consists of human, mouse, and rat) and outside this clade the exon was constitutively spliced. Thus, this gene shows the same evolutionary pattern as the *SLC35B3* gene: constitutive splicing is the ancestral state and alternative splicing is the derived one.

In the three other genes tested (*APGL5*, *SLC25A*, and *DPM1*), alternative splicing was evident in a subset of the tested mammals. The human and mouse exons were alternatively spliced in all cases, rat exons were always constitutively spliced, and the dog exons were alternatively spliced in the case of the *APGL5* gene and constitutively spliced in the others. In all cases, the splicing in nonmammalian vertebrates was constitutive, suggesting that constitutive splicing is also the ancestral form for these three genes. The observed splicing pattern of the *APGL5* gene in rat is due to (i) the presence of an alternatively spliced isoform in tissues other than the brain or under only certain environmental conditions or (ii) a reversal in rat to the ancestral constitutive state. Supporting the latter hypothesis of reversal is the observation that rodent genomes are highly dynamic [31]. In the *SLC25A* and *DPM1* genes, alternative cassette exons were found only in human and mouse. The alternative splicing in rat either was not detected or has been lost. It may be that two independent shifts from constitutive to alternative splicing have occurred: one in the lineage leading to human and one in the lineage leading to mouse after the rat–mouse divergence. The ten exons evaluated (Figure 2) have shifted their mode of splicing from constitutive to alternative during evolution. This is consistent with the high level of similarity found between species-specific AS exons and constitutively spliced exons (Figure 1), which presumably reflects their evolutionary history.

## Alternatively Spliced Exons with Higher Inclusion Levels Are More Similar to Constitutive Exons than Those with Lower Inclusion Levels

The bioinformatics analysis and the experimental assay suggest that some alternative cassette exons with a high inclusion level originated from constitutively spliced exons. Thus, we expected that alternative cassette exons with a high inclusion level that are conserved in human and mouse are under a different selective pressure than alternative cassette exons with a low inclusion level. To test this hypothesis, two groups of conserved alternative cassette exons were defined, according to their inclusion level. The high and low inclusion groups (HI and LI, respectively) included the upper and lower inclusion level quartiles, respectively (inclusion level > 93.8% and < 37.5%, respectively). We characterized these two groups in terms of their human–mouse conservation level, fraction of symmetrical exons, 5′ss score (presented by percent and calculated by Senapathy algorithm [32]), exon length, Ka/Ks likelihood ratio test (Fail Ka/Ks), Ka/Ks value, and conservation level of flanking introns. We compared these values with those in conserved constitutively spliced exons and alternative cassette exons. A significant difference was found between the LI and HI groups in all tested parameters (Figure 3; see Table S3 for the statistical analysis). In addition, for all seven parameters tested, the average value of the HI group was found to be more similar to constitutively spliced exons than those of the LI group. For example, examination of the conservation level between human and mouse revealed that constitutively spliced exons exhibited the lowest conservation level (87.8%). The HI group exhibited a conservation level of 89.8%, lower than that of the LI group (94.2%; see Figure 3A). Thus, the HI group was more similar to constitutively spliced exons than to the LI group. The alternative cassette exons showed intermediate conservation levels, as expected. The same general pattern was found with respect to the other six parameters (i.e., the HI exons were more similar to constitutively spliced exons than were other alternative cassette exons, and the LI exons were less similar to constitutively spliced exons than other alternative cassette exons were (Figure 3B–3G)). These results are consistent with other results [24,33]. This analysis suggests that the selection pressure acting on the LI and the HI groups is different, with the latter being more similar to the constitutive spliced exons group. This provides additional support for the hypothesis that some HI exons have evolved from constitutively spliced exons.

## Fixation of Exonic Splicing Regulatory Sequences

A comparative analysis of unicellular and multicellular eukaryotic 5′ss has revealed important differences—the plasticity of the 5′ss of multicellular eukaryotes means that these sites can be used in both constitutive and alternative splicing and for the regulation of the inclusion/skipping ratio in alternative splicing. So, alternative splicing might have originated as a result of relaxation of the 5′ss recognition in organisms that originally could support only constitutive splicing [6]. To examine the molecular evolutionary changes required to shift exons from constitutive splicing to alternative splicing, we analyzed the sequence of exon 5 of the human *SLC35B3* gene, which is alternatively spliced in mammals and constitutively spliced in other vertebrates (Figure 2B). A multiple sequence alignment of the 5′ss region of this exon from 11 vertebrates was constructed using MAFFT [34] (Figure 4A). The 5′ss of all seven mammals is characterized by GT at positions 6 and 7. This is not typical of mammalian 5′ss: usually GT is found at positions 5 and 6 [6]. However, in all other vertebrates, there is a GT at positions 5 and 6 (AT replaces GT in zebrafish). This finding suggests two possible scenarios: either the ancestral vertebrate had GT at positions 5 and 6 or the ancestral vertebrate had GT at positions 6 and 7. These scenarios correspond to an insertion event in the lineage leading to mammals or to two deletion events: one in the lineage leading to chicken and one in the lineage leading to fish (Figure 4A). The deletion scenario is less parsimonious and, hence, we speculate that an insertion event in a mammalian ancestor was responsible for the differential positioning of the GT in the 5′ss. Can this difference in GT location explain the observation that this exon is alternatively spliced in mammals and constitutively spliced in other vertebrate? This hypothesis was tested experimentally as described below.

To examine if the molecular changes in the 5′ss and the high level of exon conservation among conserved alternative exons are correlated, we analyzed the distribution of putative ESRs in the sequence of this exon. We aligned this exon from 15 organisms, and identified 12 putative ESRs using the ESEfinder [35]. Five were selected for further analysis (see also

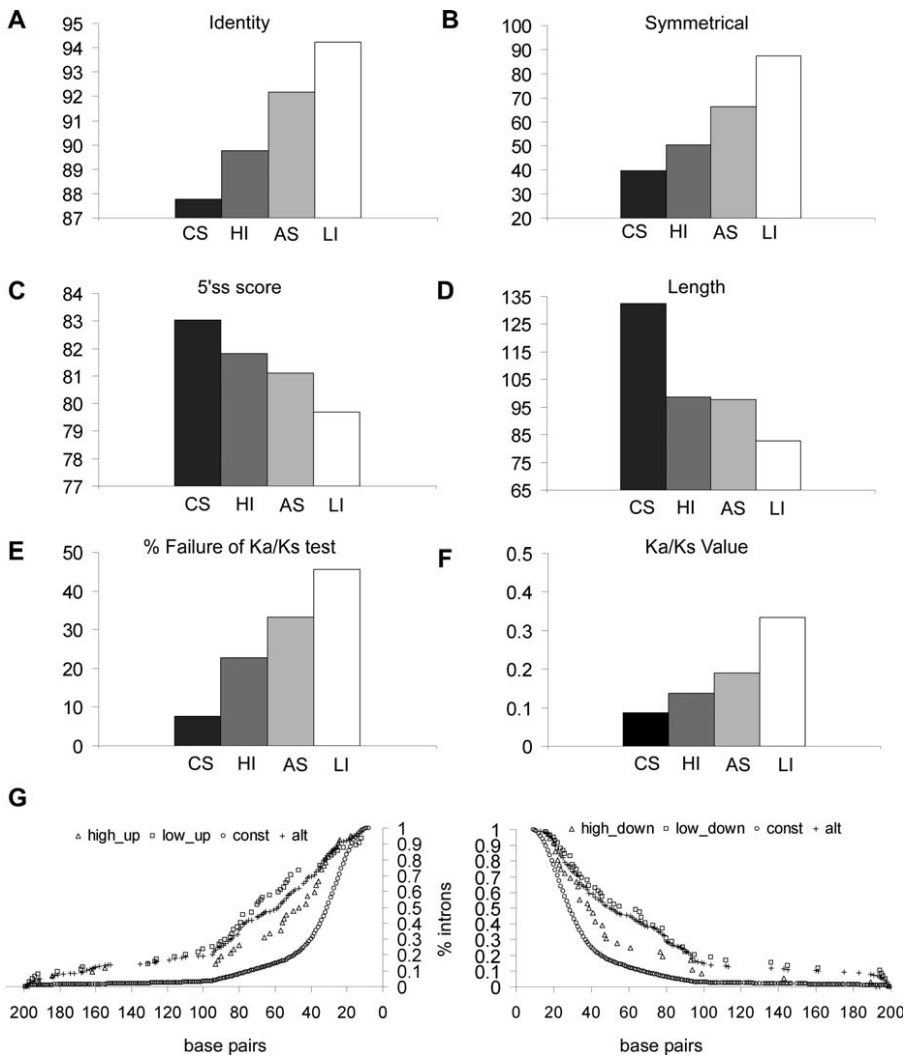

**Figure 3.** High Inclusion (HI) and Low Inclusion (LI) Alternative Cassette Exons Exhibit Different Characteristics

The conserved alternative cassette exons were tested for their inclusion level. Conserved alternative cassette exons with LI and HI were compared with conserved constitutively spliced and all alternative cassette exons: (A) exon conservation level; (B) percentage of symmetrical exons; (C) 5′ss score; (D) average length; (E) percent of exons that failed the Ka/Ks likelihood ratio test; (F) Ka/Ks value; and (G) conservation of the flanking intronic sequences. Median conservation level at each base position was calculated for the 200 base pairs upstream (left) and downstream (right) for the CS (circle), HI (triangle), AS (plus), and LI (square) exons.

doi:10.1371/journal.pgen.0030203.g003

Tables S4 and S5). The differences in ESR sequences of chicken and zebrafish relative to mammals considerably change the ESR score and are, therefore, predicted to affect splicing (these substitutions are in bold in Figure 4B; see Tables S4 and S5). ESEfinder is based on the binding score of human SR proteins [35]. However, it is expected that these ESRs bind SR proteins outside the mammalian class as well [36,37]. Our analysis revealed dynamic changes in both the 5′ss and the five different potential ESRs among the nonmammalian species and fixation of those sites among all mammals. These changes are correlated with the transition to alternative splicing.

To test the above observations experimentally, we cloned a minigene containing the *SLC35B3* human genomic sequence from exon 4 through exon 6 and transfected it into human 293T cells. Total RNA was collected, followed by RT-PCR analysis and sequencing. As shown in Figure 4C (Human panel, lane 1 named h5′ss), the exogenous wild-type (wt) mRNA was alternatively spliced with a low inclusion level. The difference in the inclusion level between the minigene and the endogenous wild type (WT) mRNA (Figure 2B), might be due to promoter differences [38]. Reconstitution of the human 5′ss to that of chicken (named c5′ss) by a deletion of T at position 4 of the human 5′ss shifted splicing from alternative to constitutive (Figure 4C, Human panel, lanes 1 and 2). This indicates that the evolutionary events differentiating the chicken and the human 5′ss are directly linked to the shift from constitutive to alternative splicing. We focused our attention on the molecular changes in the 5′ss and not the PPT/3′ss region because no change was observed in that region between constitutively and alternative cassette exons, whereas changes were observed in the 5′ss (see Text S1).

The 5′ss of zebrafish has the weakest site among the

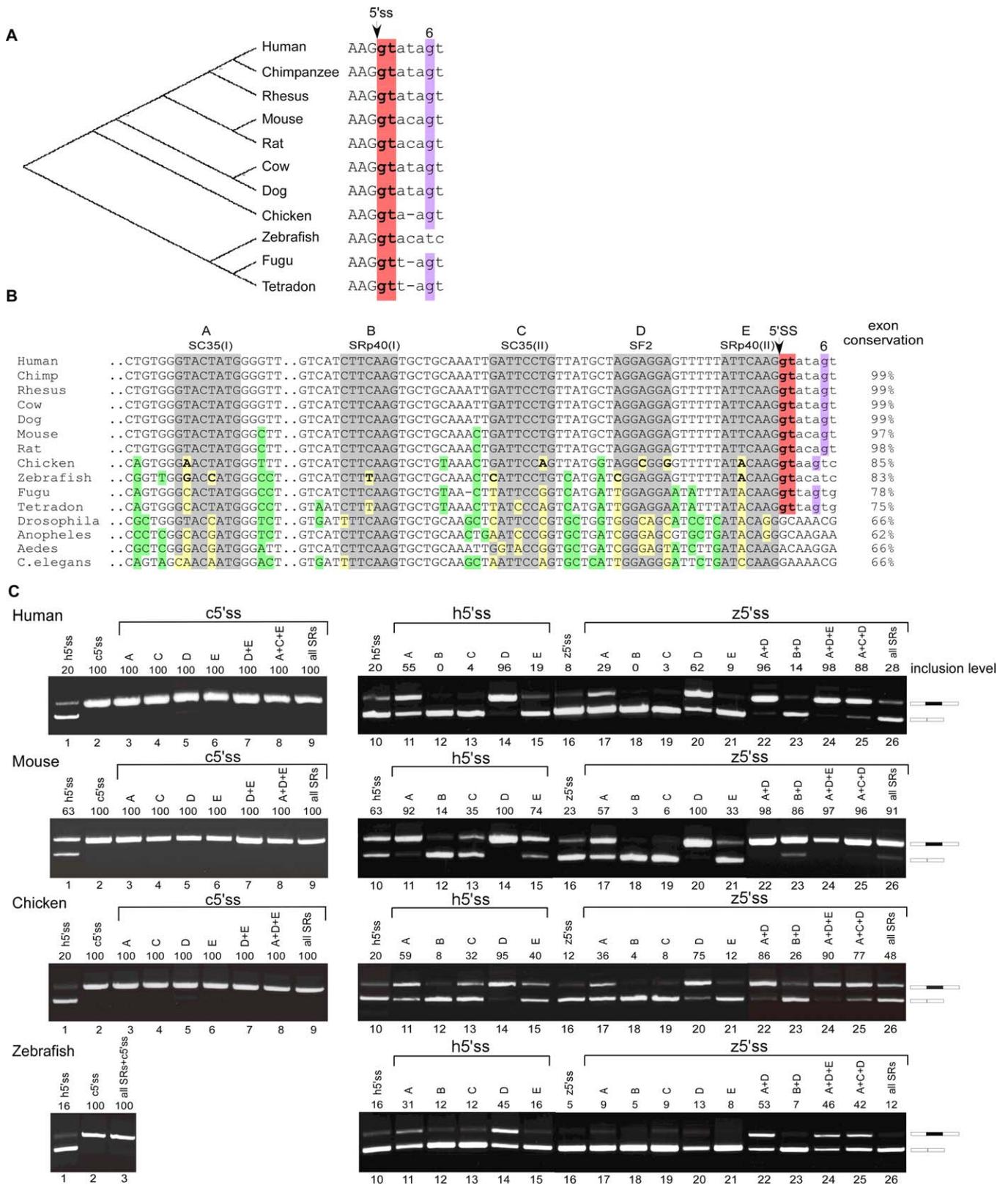

**Figure 4.** Mutations That Weaken the 5'ss during Evolution Shift Splicing from Constitutive to Alternative

(A) Multiple alignment of the 5'ss of exon 6 of the *SLC35B3* gene from 11 vertebrates was constructed, substitutions were analyzed based on the known evolutionary tree [44]. The first two intronic nucleotides are highlighted in red and evolutionary shift of the G from position 5 to 6 at the 5'ss is highlighted in purple. The gap indicates a potential insertion event. (B) Alignment of exon 5 of the *SLC35B3* gene among 15 different organisms. The first two nucleotides of intron 5 are highlighted in red (the last four species do not contain this intron). As above, the shift of the G from position 5 to 6 among those organisms in the 5'ss is highlighted in purple. The examined five ESRs are highlighted in gray and marked A to E, with the name of the putative SR protein that binds that sequence noted in the upper part of the panel. Nucleotides that are highlighted in yellow and green show exonic organisms evaluated, as it lacks the conserved G found at either position 6 in mammals or at position 5 in other vertebrates (Figure 4A). Thus, it was surprising that this zebrafish exon is constitutively spliced (Figure 2B, SLC35B3, lane Z). The constitutive splicing pattern in zebrafish presumably results from the effects of ESRs rather than the signal in the 5′ss (see also [37]). Indeed when the 5′ss of human was mutated to that of zebrafish (named z5′ss), it resulted in almost total use of the alternative cassette exon (Figure 4C, Human panel, compare lanes 10 and 16), indicating that the zebrafish 5′ss is not sufficient to generate a constitutive splicing pattern.

We next studied the combined effect of 5′ss and ESRs. Mutating the 5′ss of human to that of chicken and reconstructing ESRs A, C, D, and E to the corresponding chicken ESRs had no effect on splicing (Figure 4C, Human panel, lanes 3–6). This indicates that a strong 5′ss is dominant over the effect of ESRs. However, when the 5′ss is weak, ESR mutations considerably altered splicing. This was shown by mutations that reconstructed ESRs A–E to the zebrafish sequences in the presence of either the zebrafish 5′ss (the weakest 5′ss) or the mammalian 5′ss (intermediate 5′ss strength). When the ESRs were identical to human, the alternative splicing was observed for both the human 5′ss and the zebrafish 5′ss and the level of exon inclusion was consistent with the 5′ss strength (Human panel, compare lanes 10 and 16, respectively). Mutating the human ESRs to those found in zebrafish under both splice sites indicated that ESRs B and C function as enhancers, ESRs A and D as silencers, and that ESR E had no effect (Figure 4C, Human, lanes 10–21). This shows that ESRs A–D influence both the human and the zebrafish 5′ss.

To examine combinatorial effects of ESRs, two and three ESRs were mutated simultaneously, using either chicken or zebrafish 5′ss. The strong 5′ss of chicken was dominant over combinatorial changes in the ESRs (Figure 4C, Human, lanes 7–9). However, when the 5′ss was that of zebrafish, ESR mutations affected splicing considerably: mutations in ESRs A+D, A+D+E, and A+C+D enhanced the inclusion level while mutations in B+D and A+B+C+D+E had a minor effect on the inclusion level (Figure 4C, Human, lanes 22–26). The sum of the scores according to the ESEfinder was predictive of the changes in inclusion level observed by RT-PCR experiments (see Table S4 B). This result implies a delicate combinatorial effect among the ESRs on the splicing of an exon.

The high level of ESR conservation among mammals implies that those sites are essential for the regulation of the inclusion/skipping level only when the exon is alternatively spliced (Figure 4B–4C). However, the low conservation level of those sequences among the nonmammalian exons might indicate that ESRs are not essential when an exon is constitutively spliced or that the ESRs are not functional splicing regulatory elements in nonmammalians. To examine this, we transfected the WT and mutated plasmids to mouse and chicken cell lines (NIH 3T3, DF-1, respectively) and to zebrafish embryos (Figure 4C marked Mouse, Chicken, and Zebrafish). In general, splicing patterns were similar to that of human for the three types of 5′ss and for all mutated ESRs. For example, under the strong chicken 5′ss, a constitutive pattern was observed in all species (Mouse and Chicken panels lanes 2–9, Zebrafish panel lanes 2–3). Similarly, ESRs B and D acted as a strong enhancer and a strong silencer, respectively, in all species (see all panels lanes 14 and 20 for ESR D and lanes 12 and 18 for ESR B). These findings suggest that these ESRs are active and have a similar function in all tested cells. This further strengthens the notion that the differences in the inclusion levels between the mouse and the other species do not reflect divergent evolution of splicing factors, but, rather, a different cellular environmental effect. We note that in zebrafish the effect of ESRs was generally weak compared with the other organisms (compare lane 14 in all panels, see also Table S4). These results imply that fixation of the ESRs is linked to the shift from constitutive to alternative splicing.

## The Link between a Weak 5′ss and Exonic Splicing Enhancers and Silencers

To further characterize the evolutionary processes that are linked with the transition from a strong to a weak 5′ss, the species-specific exons were subdivided into two groups: those in which the 5′ss changed (54%, 330 out of 612 species-specific exons) and those in which the 5′ss did not change (46%, 282 out of these 612 exons) between human and mouse. For those exons in which the 5′ss changed, a significant decrease in 5′ss strength was observed, from 83.34% to 82.02%, between constitutively and alternative cassette exons, respectively (Mann-Whitney, $p$ value < 0.05). Based on the information content equation (also described in [39]) the 5′ss information content decreased from 8.25 to 7.91 bits. However, in the group in which there was a complete conservation of the 5′ss sequence between constitutively and alternatively spliced pairs, the average 5′ss score was already weak (82.52%), similar to the 5′ss score of conserved alternative exons (Mann-Whitney, $p$ value = 0.125). The changes in the 5′ss region between conserved constitutively spliced exons and alternative cassette exons is shown in Text S1 and Figure S1.

The two groups were further analyzed for their Ka/Ks ratio, their Ks score, and their conservation level (Table S6). A substantially higher synonymous substitution rate was found in the group in which the 5′ss changed, compared to the group in which the 5′ss did not change ($\chi^2$, $p$ value < $1.8 \times 10^{-32}$ for Ks, $1.82 \times 10^{-6}$ for Ka, and $3.5 \times 10^{-37}$ for the Ka/Ks). In addition, the conservation level (percentage of identity) significantly differed between the two groups (Mann-Whitney, $p$ value < $2.7 \times 10^{-8}$).

To correlate these findings with creation/elimination of exonic splicing enhancer (ESE) and exonic splicing silencer

(ESS) regions, we compared the ratio between the number of ESEs and ESSs (ESE/ESS ratio) between the species-specific AS exons to the species-specific constitutive ones. Specifically, 238 RESCUE-ESEs [40] and 176 ESSs [41] were separately compared for the exons pairs in which the 5′ss changed and those in which the 5′ss did not change. For those in which the 5′ss changed, that ESE/ESS ratio was lower for the alternatively spliced exons than for the constitutively spliced ones (2.46 compared with 3.19, respectively; $\chi^2$, $p$ value = 0.0026; see Table S7). No statistically significant difference was found in the ESE/ESS ratio between the species-specific spliced alternative and constitutive exons in which the 5′ss did not change during evolution ($\chi^2$, $p$ value = 0.24). Next, the difference in the ESE/ESS ratio between the two groups of species-specific AS exons was examined. The group of species-specific AS exons in which the 5′ss changed during evolution (i.e., decrease in 5′ss affinity) showed a lower ESE/ESS ratio compared with the group of species-specific exons in which the 5′ss did not change (2.46 compared with 2.92, respectively; $\chi^2$, $p$ value = 0.028). This indicates that the evolutionary shift from constitutive to alternative splicing that is manifested by mutations that weaken the 5′ss is frequently associated with changes (mostly synonymous substitutions) in the ESE/ESS ratio in the exonic region located upstream of the 5′ss. These changes presumably ensure proper exon recognition and regulation of the skipping/inclusion level. To control for potential bias in our results due to higher conservation level of the unchanged 5′ss group compared to the changed 5′ss group, we repeated the analysis by constructing two groups with identical conservation level (by cutting off the top 15% of the conserved exons from the non-changed group and cutting the bottom 15% from the group of the conserved exons that their 5′ss was changed); the results were similar to the results reported above.

## The Importance of the Branch Site in the Shift from Constitutive to Alternative Splicing

The above findings suggest that mutations that weaken the 5′ss during evolution are one of the mechanisms responsible for the shift from constitutive to alternative splicing. We hypothesize that constitutively spliced exons with a weak 5′ss may be on the verge of shifting to alternative splicing. To test this hypothesis, we searched for constitutively spliced exons with a weak 5′ss in human and a strong 5′ss in mouse or vice versa. We found 593 cases in which the human 5′ss has G at position 5 and the mouse has G at position 6 and the opposite in 725 cases. We selected one of these exons (exon 12 of *IMP* gene, which was previously used in our lab [42]) and constructed a multiple sequence alignment of the orthologous exons and their flanking intronic sequences from ten vertebrates using the M-LAGAN tool ([43], Figure 5A). The 5′ss of human, chimp, rhesus, and cow are characterized by a G at position 6, whereas in mouse and rat the G is at position 5. In dog, there is no G in positions 5 or 6 and, based on the close phylogenetic relationship between the cow and dog 5′ss [44], we suggest that a transversion from G to C in the lineage leading to dog is responsible for this change. It is possible that the 5′ss of mouse and rat evolved through a single deletion of the T at position 3 of the human 5′ss, which resulted in the shifting of the G at position 6 to 5. This shift strengthens the 5′ss from a score of 61.18 in human to 73.5 in mouse (based on

http://ast.bioinfo.tau.ac.il/SpliceSiteFrame.htm). The 5′ss could not be aligned reliably to the nonmammalian vertebrates.

We first validated that both the human and the mouse 5′ss are associated with constitutively spliced isoforms in a minigene containing exons ten through 13 of the human *IMP* gene (Figure 5B, lanes 1 and 2). We next tested the effect on splicing of a single mutation in a putative BS motif on both the human and the mouse 5′ss. This putative BS was found based on the Kol et al. algorithm that compares human and mouse genomes [42]. The putative BS and the surrounding region were mutated in five different positions [42]. We also mutated six adenosines located at the 3′ end of this intron; none of the mutations affected splicing (see Figure S4). We mutated the conserved A in the putative BS to G (Figure 5A). Under the 5′ss of human, the mutation shifted splicing from constitutive to alternative (lane 3). However, under the mouse 5′ss, this mutation did not change the splicing pattern (lane 4), indicating that a strong 5′ss was dominant over mutation at the putative BS. Similar results were obtained when a T to G mutation at position −2 of the putative BS was tested (lanes 7 and 8). A G to C mutation at position −1 of the putative BS (lanes 5 and 6, respectively) had no effect. The mutations tested in lanes 3, 5, and 7 of Figure 5A were evaluated previously by Kol et al. [41]. The importance of positions 1 and −2 of the putative BS [42] is also reflected by the conservation of these positions in all vertebrates, whereas the G at position −1 is not conserved outside mammals (Figure 5A).

Establishing that the human exon is on the verge of becoming alternatively spliced, we next tested whether mutations in the putative ESRs can also shift the splicing pattern of this exon to alternative. We previously identified two ESRs (L and M) in this exon [29]. Deleting ESR L had no effect on splicing under both the human and the mouse 5′ss (Figure 5B, lanes 9 and 11). However, when the deletion was combined with the A to G mutation in the putative BS, the inclusion level changed from 54% to 86% under the human 5′ss (lane 3 versus lane 10). This indicates that ESR L is a silencer under these conditions. However, no effect was found for these combined mutations under the strong mouse 5′ss (lane 12). Thus, ESR L affects splicing only in combination with a mutation in the putative BS motif that shifts splicing from constitutive to alternative and only with a weak 5′ss.

A point mutation in ESR M had no effect on the constitutive recognition of that exon under either the human and mouse 5′ss (lanes 13 and 15, respectively). However, when the ESR M mutation was combined with the A to G mutation in the putative BS under the human 5′ss, splicing was shifted completely from alternative to use of the alternative cassette exon (lanes 3 and 14, respectively). Therefore, under these conditions, ESR M is a strong enhancer. The same combined mutations under the mouse 5′ss resulted in a shift to alternative splicing (lane 16), indicating that the effect of the strong enhancer ESR M can overcome a strong 5′ss only when the putative BS sequence is also mutated. These results indicate that, for the tested exon, ESRs affect splicing only when the recognition of that exon is suboptimal. In this case, suboptimization is achieved by an interplay between a mutated putative BS motif and the naturally weak 5′ss that flanks the human alternatively spliced exon.

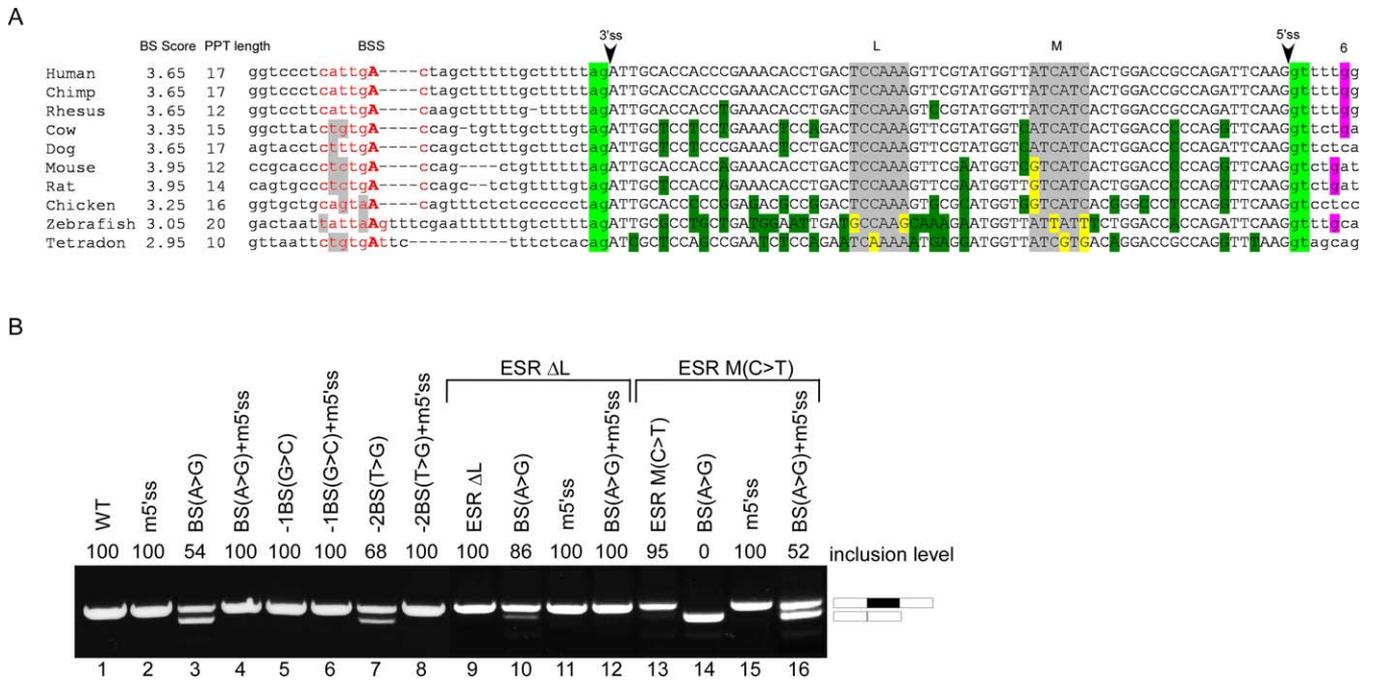

**Figure 5.** The Effects of ESRs on Constitutive and Alternative Splicing

(A) Alignment of exon 12 of the IMP gene among 10 different organisms. The BS score and the PPT length are indicated on the left. The putative BS is in red, with the branch-point A in uppercase. BS sites that differ from the human sequence are highlighted in gray, with the mutated nucleotides in bold. The AG and GT of the 3′ and 5′ss are highlighted in green. The examined ESR sites L and M are highlighted in gray, ESR L was deleted and ESR M was disrupted by a C to T mutation. The shift of the G from position 5 to 6 in the 5′ss is highlighted in purple. Nucleotides that are highlighted in yellow and green show exonic positions that were changed during evolution (with respect to the human sequence) inside the five ESRs and outside of those sites, respectively. Upper and lower case indicate exonic and intronic sequence, respectively. (B) WT and mutant plasmids were introduced into 293T cells by transfection. Total RNA was extracted and splicing products were separated on a 2% agarose gel after RT-PCR analysis. Lane 1, splicing of the WT IMP minigene. Lanes 2 through 16, splicing products of the indicated mutants. The following abbreviations are used: m5′ss indicates mutations that reconstitute the human 5′ss to that of mouse, 1BS(G>C) indicates mutation at position −1 relative to the branch-point nucleotide from G to C, ESR ΔL and ESR M(C>T) indicate deletion of ESR L and a point mutation from C to T in ESR M, respectively. Numbers on top indicate percentage of exon inclusion as determined by ImageJ analysis. The two minigene mRNA products are shown on the right.

doi:10.1371/journal.pgen.0030203.g005

## Discussion

Two models had been described for the origin of alternative cassette exons—exonization of intronic sequences and exon shuffling—that describe how new exons are added to existing genes. Our results suggest a third pathway for the evolution of alternative cassette exons: the relaxation of constitutively spliced exons. We also suggest a molecular mechanism for this transition: mutations weaken the splice sites along evolution resulting in a switch from constitutive to alternative splicing. The species-specific exons are more similar to conserved constitutively spliced exons than to conserved alternative cassette exons. However, they exhibit a minor shift toward alternative cassette exons, suggesting the shift from constitutive splicing to alternative splicing starts to impose selective pressures, manifested in such characteristics as the 5′ss motif strength, length, etc. Given the appropriate evolutionary period, these exons may either revert to the constitutive splicing pattern (as the rat exon in APGL5, Figure 2B), or further progress toward alternative splicing characteristics [24,45,46].

What is the contribution of the evolutionarily new isoforms (alternative cassette exons) to proteome diversity? For 120 species-specific spliced exons, we found indications in protein databases for the expression of the exon-inclusion isoform. Both exon included and exon skipped isoforms were found for 25 out of these 120 exons; all 25 were symmetrical exons (Text S3; S2). This suggests that exons that recently shifted to the mode of alternative splicing have the potential to contribute to protein functionality. Some of these new isoforms have been characterized: One of these is Smac3 a novel Smac/DIABLO splicing variant, that was shown to attenuate the stability and apoptosis-inhibiting activity of X-linked inhibitor of apoptosis protein [47]. Another example is the AAAS-V2, a novel splice variant of the human AAAS gene [48]. These examples indicate that some of the new isoforms resulting from skipping of alternative cassette exons function in cells.

Do species-specific AS exons reflect splicing errors? Probably some of these exons are indeed splicing errors, reflecting the low fidelity of the spliceosome machinery [49,50]. However, high inclusion level is not indicative that the alternative cassette exons isoform is a nonfunctional product, for example in SLC35B3 (Figure 2B), the low skipping level is conserved in all mammals tested. This is presumably indicative of functionality of the isoform without the exon, or else the skipped isoform would not be conserved in the mammalian class, as negative selection would act against these events and eliminate them from the species transcriptome. It is likely that the shift from constitutive to alternative splicing will at first result in high inclusion levels of the exon. With time, positive selection may increase the fraction of the isoform without the exon within the transcriptome [8,15,18]. These newcomers, the high-inclusion exons, are the playground for future exaptation, namely, the acquisition of a novel functionality different from their original [45] and fixation within the human transcriptome. Long evolutionary periods are needed for successful exaptation events, and these new alternatively used exons are the potential raw material for future evolution [15,18,45,46].

To further validate this evolutionary hypothesis, we experimentally evaluated the splicing patterns of orthologous exons in several vertebrates. We found that in the five tested exons that are alternatively spliced in human but constitutively spliced in mouse, the corresponding exons in all other tested organisms were constitutively spliced. The most parsimonious explanation of these data is that these human-specific alternative cassette exons originated from constitutively spliced exons. Since major shifts in the inclusion level are probably deleterious, the newly alternatively spliced exon should exhibit a high inclusion level. Indeed, we found high inclusion levels in all exons tested, consistent with previous studies [10].

Furthermore, the issue here is whether there exists an evolutionary force that progressively converts constitutive to alternative splicing during evolution. We do not believe that this is a progressive force, because a similar number of alternative exons in mice are constitutive in humans. This suggests that the transition from constitutive to alternative splicing is a random process, in which mutations that accumulate during evolution (and other changes in the genome) shift exons from constitutive to alternative. The fixation of the alternative form is then subjected to purifying selection. In some diseases, for example, mutations that shift splicing of exons from constitutive to alternative occur, however this shift causes deleterious effects due to the reduction of the ancestral mRNA below a certain threshold or due to the synthesis of a dominant negative protein, and thus the alternative form should eventually be purified from the population. An example is Familial Dysautonomia, in which a point mutation from T to C at position 6 of the 5′ss shifts a constitutively spliced exon to alternative splicing [51]. We anticipate that shifts in splicing are more common in organisms containing longer introns (>500 nucleotides), because alternatively spliced exons are generally flanked by longer introns than are constitutive ones [4]. Therefore, a mutation that affects exon selection in exons flanked by longer introns might be prone to shift the exon to alternative splicing.

The two known models for the origin of alternative cassette exons—exonization of intronic sequences and exon shuffling—describe how new exons are added to existing genes. Our results suggest a third pathway for the evolution of alternative cassette exons: the relaxation of constitutive splicing. We found that inclusion levels indicate the most plausible evolutionary scenario for the origin of alternative cassette exons. This is based on our bioinformatic observation that alternative cassette exons with higher inclusion levels are more similar to constitutively spliced exons than are those with low inclusion levels.

It is currently unknown whether orthologous splicing regulatory proteins from different organisms affect splicing in a similar fashion in all organisms. A recent study revealed conservation of the protein domain structure of SR proteins across all vertebrates [52]. Moreover, recent analysis implied that a *Drosophila* SR protein can also bind the corresponding human ESR domain [36]. The effect of individual ESRs, as well as their similar combinatorial effects on splicing, appears to be conserved among vertebrates (Figure 4). This suggests an important functional role of these ESRs in regulating splicing patterns and, presumably, also a conserved functional role for the proteins they bind, despite the ~450 million years of evolution since the last common vertebrate ancestor. The evolutionary insertion event in the 5′ss that resulted in weakening of the mammalian 5′ss of exon 5 of *SLC35B3* is the main driving force in the shift from constitutive to alternative splicing of this exon. Following this 5′ss weakening, the ESRs became important; a strong 5′ss is not affected by the ESRs (see also [53]). The gain of functionality of the ESRs following the shift to a suboptimal 5′ss explains the extreme increase in ESR conservation among mammals and presumably explains the higher conservation level of human–mouse orthologous exons that are alternatively spliced in both species compared to those that are constitutively spliced. The ESRs are important for exon definition and proper regulation of the inclusion level of alternatively spliced exons. Proper recognition of constitutively spliced exons, on the other hand, is ensured by stronger flanking splice sites, shorter flanking introns, and longer exons than found in alternatively spliced exon regions.

Here we demonstrated a new evolutionary pathway by which exons shifted from constitutive to alternative splicing. Exons that shift from constitutive to alternative splicing generate new isoforms and become exapted. This provides a simple mechanism to enhance transcriptomic diversification within well-established genes, and thus plays an important role in speciation. Although our major analysis was performed on human–mouse ortholog exons that diverged about 90 million years ago, we suggest that the constitutive to alternative splicing shift is an ancient mechanism that governed the formation of alternative cassette exons in the early stages of eukaryotic evolution.

## Materials and Methods

**Dataset compilation.** A dataset of 596 alternative cassette exons, conserved between human and mouse, was derived from a previously compiled dataset [5]. In addition, 45,553 human–mouse conserved constitutively spliced exons were obtained from Carmel et al. [21]. Species-specific exons were extracted from a dataset of 4,262 human–mouse orthologous exons that are suspected to splice differentially between human and mouse based on initial EST analysis [21]. However, not all these 4,262 exons necessarily reflected genuine species-specific splicing cases. This might be due to rare abnormal splicing. Thus, alternative cassette exons for which the inclusion level was higher than 98% were removed. In addition, conserved alternative cassette exons may be misclassified as species specific. In such cases an alternative exon is misclassified as constitutive due to insufficient coverage of ESTs. In other words, an exon should be defined as constitutive only if a sufficient number of ESTs reliably indicate that the exon is not alternatively spliced. The number of ESTs needed depends on the inclusion level (see [21] for how inclusion levels are computed). Modrek and Lee have shown a strong correlation between the inclusion level of homologous alternative cassette exons between human and mouse [8]. We assumed that if alternative splicing was conserved for homologous exons, we would find the same inclusion level for both species. Based on this assumption, we computed the expected number of alternative cassette exons for each number of ESTs. If three or more ESTs were expected and none were observed, we rejected the hypothesis that both exons were alternatively spliced. For example, if the alternative exon had an inclusion level of 90%, we would conclude that splicing was constitutive only if none of 30 ESTs were alternatively spliced. This is because with 30 ESTs we would expect three to support skipping. Following these filtering steps, 612 species-specific spliced exons remained. All datasets are available on line (http://www.tau.ac.il/~gilast/sup__mat.htm).

**Ka/Ks ratio test.** In order to calculate the number of exons that passed the Ka/Ks ratio test, we implemented the procedure of Nekrutenko et al. [26]. We ran the codeml program from the PAML package [54] on each exon alignment twice. First, with the Ka/Ks ratio fixed at 1 followed by analysis with the Ka/Ks ratio as a free parameter. We then collected the maximum likelihood values ML1 and ML2 from the two runs and calculated the likelihood ratio as $LR = 2(lnML1 - lnML2)$. Next, we compared the $LR$ against the $\chi^2$ distribution with one degree of freedom to test whether Ka/Ks was significantly different from 1 [55].

**Calculation of the flanking introns conservation level.** Intron alignment between human and mouse were calculated according to the UCSC pairwise chained blastz alignments found in <chrN.hg18.mm8.net.axt> files, based on the hg18 (Mars 2006) and mm8 (Mars 2006) versions of the human and mouse genomes, respectively. These files were downloaded from the UCSC genome browser (http://hgdownload.cse.ucsc.edu/goldenPath/hg18/vsMm8/). From these files the length of the intronic alignments between human and mouse were retrieved.

**Defining high and low inclusion level of alternative exons.** The conserved alternative cassette exons were divided into subgroups according to their inclusion level. The 25% of exons with the lowest and highest inclusion levels were defined as low inclusion (LI) and high inclusion (HI) alternative cassette exons, respectively. These definitions resulted in 149 HI exons with inclusion levels higher than 93.8% and 151 LI exons with inclusion levels lower than 37.5%.

**Statistical analysis.** The calculation of $p$-value was measured according to the data distribution. The Kolmogorov-Smirnov test was used to examine normal distribution. The T-test was used to calculate statistical differences for $p \geq 0.05$; otherwise ($p$ value < 0.05), the Mann-Whitney test was used for $p$-value calculations.

**cDNA samples.** We used commercial brain cDNA from human, mouse, rat, and dog (BioChain) and prepared cDNA from different cell lines, including human U2OS (bone osteosarcoma cells), mouse NIH 3T3 (embryonic fibroblast cells), mouse DA-3 (mammary carcinoma cells), and rat H4-II-E-C3 (hepatoma cells). Chicken (5-d-old embryos) and zebrafish (adults) were disrupted in TRIzol (Sigma) with a Polytron homogenizer (PT-MR2100 Kinematica AG, Switzerland). Xenopus oocytes were disrupted with a hand-held motor-pestle (kimble-kontes, NJ). After the complete homogenization of the tissue, total RNA extraction and reverse-transcription reaction were performed as described in below.

**Primer construction for endogenous amplification.** Species-specific primer pairs were designed using the UCSC Genome Browser [56] to target the immediately flanking constitutive exon sequences and to amplify an identical sized PCR product (of the exon inclusion form) from different species. Oligonucleotide sequences will be provided upon request.

**Transient expression assay in vivo.** Transient expression assays of the pEGFP-containing constructs were performed by microinjection of zebrafish embryos as previously described [57,58]. WT and mutant plasmids were diluted to a final concentration of 100 ng/µl injection solution (0.1 M KCl, 0.05% phenol red). Approximately 2 ng of plasmid DNA were injected into the cytoplasm of one- or two-cell stage zebrafish zygotes using a micromanipulator and a PV830 Pneumatic Pico Pump (World Precision Instruments, Sarasota, FL, USA). Embryos (~300) were injected with three to four different needles for each construct. The embryos were then placed in 10 cm Petri dishes with egg water [59] containing methylene blue (0.3 ppm) and raised in a light- and temperature-incubated incubator (28 °C; light intensity 12 W/m²). Green fluorescence in live embryos was detected at 1 day post fertilization under an Olympus dissecting microscope SZX12 equipped with EGFP filter for excitation (460–490 nm) and emission (510–550 nm) (see Figure S3). pEGFP-positive embryos were collected, frozen, and stored in liquid nitrogen prior to RNA extraction.

**Plasmid constructs.** The desired minigene was generated by amplifying a human genomic fragment using PCR (Text S4). Each primer contained an additional sequence encoding a restriction enzyme. The PCR products were restriction digested and inserted into the pEGFP-C3 plasmid (Clontech), subsequently the minigene was sequenced. The first cloned minigene was from the *SLC35B3* gene (solute carrier family 35 member B3) containing exons 4 through 6 (3.1 kb). Due to size constraints that dictate what can be inserted into this vector, we shortened the cloned fragment by deletion of 4,350 nucleotides from the middle of intron 4. This was done by amplifying two PCR products and subcloning the fragments. The second minigene was from the *IMP* gene (IGF-II mRNA-binding protein) and contained exons 10 through 13 (2.6 kb).

**Site-directed mutagenesis.** Oligonucleotide primers containing the desired mutations were used to amplify the mutation-containing replica of the wild-type minigene plasmid. The products were treated with DpnI restriction enzyme (12U, New England BioLabs) at 37°C for 1h. From 1 to 4 µl of the mutant DNA was transformed into *E. coli* XL1-competent cells. Colonies were picked and DNA purified by Midi-prep (GENOMED). Sequences of all plasmids were confirmed by sequencing.

**Transfection, RNA isolation, and RT-PCR amplification.** 293T, NIH 3T3, DA-3, and DF-1 cell lines were cultured in Dulbecco's Modification of Eagle Medium, supplemented with 4.5 g/ml glucose (Biological Industries Inc., Israel), 10% fetal calf serum (FCS), and 100 U/ml penicillin, 0.1 mg/ml streptomycin, and 1 U/ml nystatin (Biological Industries Inc., Israel). For the DF-1 cell-line, 1.5 g/l sodium bicarbonate was also added. Cells were cultured in 60 mm dishes under standard conditions at 37 °C with 5% $CO_2$. Cells were grown to 60% confluence and transfection was performed using 3 µl FuGENE6 (Roche) with 1 µg of plasmid DNA. After 48 h cells were harvested. Total RNA was extracted using TRIzol Reagent (Sigma). The injected zebrafish embryos were disrupted in TRIzol (Sigma) using a hand-held motor-pestle (kimble-kontes, NJ). All samples were treated with 2 U of RNase-free DNase (Ambion). Reverse transcription was performed on 1–2 µg total RNA using reverse transcribed avian myeloblastosis virus (RT-AMV, Roche) following the manufacturer's protocol. The spliced cDNA products derived from the expressed minigenes were detected by PCR using Taq polymerase (BioTools) and pEGFP-C3-specific reverse and forward primers. Amplification was performed for 30 cycles, consisting of denaturation for 30 s at 94 °C, annealing for 45 s at 58 °C, and elongation for 1–2 minutes at 72 °C. The products were separated in a 2% agarose gel. Although the RT-PCR analysis is a semi-quantitative method, the differences between the conditions tested here and a quantitative RT-PCR analysis are typically no more than 10% [60].

Additional methods are found in Text S6.

## Supporting Information

**Figure S1.** The Major Changes in the 5'ss between Conserved Constitutive, Alternative, and Species-Specific Exons

Found at doi:10.1371/journal.pgen.0030203.sg001 (137 KB DOC).

**Figure S2.** Symmetrical Exons Position within the Coding Sequence (CDS) Shows Equal Distribution along the CDS, Whereas Showing a Preference for Non-symmetrical Exons to Located at the 5' Half of the CDS

Found at doi:10.1371/journal.pgen.0030203.sg002 (23 KB DOC).

**Figure S3.** CMV Promoter-Driven Transient Expression of WT *SLC35B3* Minigene in Zebrafish Embryo

Found at doi:10.1371/journal.pgen.0030203.sg003 (2.9 MB DOC).

**Figure S4.** Splicing Assays on Putative BS Mutants

Found at doi:10.1371/journal.pgen.0030203.sg004 (100 KB DOC).

**Table S1.** $p$-Values Attributed to the Different Characteristics between Constitutive, Alternative, and Species-Specific Exons

Found at doi:10.1371/journal.pgen.0030203.st001 (55 KB DOC).

**Table S2.** Accession Numbers for the Genes Used in This Manuscript.

Found at doi:10.1371/journal.pgen.0030203.st002 (33 KB DOC).

**Table S3.** The Differences between Low Inclusion and High Inclusion Alternative Exon Groups as Described Graphically in Figure 3

Found at doi:10.1371/journal.pgen.0030203.st003 (40 KB DOC).

**Table S4.** The Five Putative SR Binding Sites and Their Different Scores among Species

Found at doi:10.1371/journal.pgen.0030203.st004 (44 KB DOC).

**Table S5.** List of Potential SR Binding Sites That Were Found in Exon 5 of the Human *SLC35B* Gene

Found at doi:10.1371/journal.pgen.0030203.st005 (76 KB DOC).

**Table S6.** Ka/Ks Analysis of Species Specific

Found at doi:10.1371/journal.pgen.0030203.st006 (34 KB DOC).

**Table S7.** Two Different Groups of Alternative 5′ss and Their ESE/ESS Ratio

Found at doi:10.1371/journal.pgen.0030203.st007 (27 KB DOC).

**Text S1.** Comparative Analysis of 5′ss

Found at doi:10.1371/journal.pgen.0030203.sd001 (28 KB DOC).

**Text S2.** Figure 4 Additional Explanation

Found at doi:10.1371/journal.pgen.0030203.sd002 (24 KB DOC).

**Text S3.** The Functionality of the Exon Skipping Isoforms of the Species-Specific Spliced Exons

Found at doi:10.1371/journal.pgen.0030203.sd003 (25 KB DOC).

**Text S4.** Minigene Sequences.

Found at doi:10.1371/journal.pgen.0030203.sd004 (32 KB DOC).

**Text S5.** Description of the Non-conserved Spliced Isoforms from Figure 2

Found at doi:10.1371/journal.pgen.0030203.sd005 (26 KB DOC).

**Text S6.** Additional Materials and Methods

Found at doi:10.1371/journal.pgen.0030203.sd006 (30 KB DOC).

## Acknowledgments

**Author contributions.** GLM, AG, NS, and GA conceived and designed the experiments. GLM, AG, NS, HK, and SLB performed the experiments. GLM, AG, NS, EK, and ADF analyzed the data. GLM, AG, NS, TP, and GA wrote the paper.

**Funding.** This work was supported by a grant from the Israeli Science Foundation (1449/04 and 40/05), MOP Germany-Israel, GIF, DIP, and EURASNET. TP was supported by an Israeli Science Foundation grant number 1208/04 and by a grant from the Israeli Ministry of Technology.

**Competing interests.** The authors have declared that no competing interests exist.